\documentclass[twocolumn,showpacs,preprintnumbers,amsmath,amssymb,floatfix]{revtex4-1}

\usepackage{latexsym}
\usepackage{amsmath}
\usepackage{epsfig}
\input{epsf.sty}  

\usepackage{epstopdf}
\usepackage{graphicx}
\usepackage{dcolumn}
\usepackage{bm}
\usepackage{subfigure}
\usepackage{enumerate}

\begin{document}

\preprint{version/03-09-16}

\title{Obtaining mass parameters of compact objects from red-blue shifts 
emitted by geodesic particles around them.}

\author{
Ricardo Becerril$^{1}$,
Susana Valdez-Alvarado$^{2}$ and
Ulises Nucamendi$^{1}$,
}

\affiliation{
$^{1}$Instituto de F\'isica y Matem\'aticas, Universidad
Michoacana de San Nicol\'as de Hidalgo. Edif. C-3, 58040 Morelia, 
Michoac\'an, M\'exico, \\
$^{2}$Facultad de Ciencias de la Universidad Aut\'onoma del Estado de
M\'exico, Instituto Literario No. 100, C.P. 50000, Toluca, Estado de M\'exico, M\'exico.}
\date{\today}

\begin{abstract}

The mass parameters of compact objects such as Boson Stars, 
Schwarzschild, Reissner Nordstrom and Kerr black holes are computed 
in terms of the measurable redshift-blueshift ($z_{red},z_{blue}$) 
of photons emitted 
by particles moving along circular geodesics around these objects 
and the radius 
of their orbits. We found bounds for the values of 
($z_{red},z_{blue}$) that may be observed. For the case of Kerr
black hole, recent observational estimates of 
SrgA$^*$ mass and rotation parameter 
are employed to determine the corresponding values of these
red-blue shifts.
 
\end{abstract}

\maketitle

\section{Introduction}\label{sec:int}

The increasing amount of evidence that many galaxies contain 
a supermassive black hole at their center \cite{evidence}, motivated
Herrera and Nucamendi (hereafter referred as H-N) to develop a theoretical 
approach to obtain 
the mass and rotation parameter of a Kerr black hole in terms of the 
redshift $z_{red}$ and 
blueshift $z_{blue}$ of photons emitted by massive particles 
traveling around them along geodesics and the radius of their orbits 
\cite{ulises}. They found an explicit expression of the 
rotation parameter as a function of $z_{red}$, $z_{blue}$, the radius
of circular orbits and the mass $M$, whereas $M$ might be found by solving
an eight order polynomial which can only be done numerically. 
These circular orbits should of course, be bounded and stable. 
If a set of observational data $\{z_{red},z_{blue},r\}$, that is,
a set of red and blue shifts emitted by particles orbiting 
a Kerr black hole 
at different radii were given, 
what would be desirable to know is the
mass and rotation parameter in terms of that data set. 
In this paper, we provide the details of  
how this can be accomplished. Particularly, the mass of the black hole  
for SgrA$^{*}$ and its corresponding 
angular momentum that have been recently estimated
\cite{estimates}: $M \sim 2.72 \times 10^6 M_{\odot}$
and $a \sim 0.9939 M$, are employed in our analysis.
In addition, the mass parameter of axialsymmetric 
non-rotating compact objects such as 
Schwarzschild and Reissner-Nordstrom black holes as well as Boson-Stars
is found in terms of the red-blue shift of light and the 
orbit radius of emitting particles. In order to have a self contained 
paper, we provide a brief summary of H-N theoretical
scheme in the section II. In sections III and IV we deal with the 
non-rotating examples above mentioned and the rotating
Kerr black hole respectively.
 
\section{Theoretical Approach}
\label{review}

H-N considered a rotating axialsymmetric space-time
in spherical coordinates $(x^{\mu})=(t,r,\theta,\phi)$.
The geodesic trayectory followed by a massive particle in 
this space-time can be obtained by solving
the Euler-Lagrange equations

\begin{equation}
\frac{\partial \mathcal{L}}{\partial x^{\mu}} -
\frac{d}{d \tau} \left ( \frac{\partial \mathcal{L}}{\partial \dot{x}^{\mu}}  
\right )=0,
\label{ELE}
\end{equation}

\noindent with the Lagrangian $\mathcal{L}$ given by

\begin{equation}
\mathcal{L}= \frac{1}{2} \left [
g_{tt} \dot{t}^2+2 g_{t\phi}\dot{t} \dot{\phi} + g_{rr} \dot{r}^2 + 
g_{\theta \theta} \dot{\theta}^2+ g_{\phi \phi} \dot{\phi}^2 \right ],
\label{lagrangian}
\end{equation}

\noindent 
being $\dot{x}^{\mu}= \frac{d x^{\mu}}{d\tau}$
and $\tau$ the proper time. It is assumed that the metric depeds
solely on $r$ and $\theta$; thus, the space time is endowed 
with two commuting Killing vectors $[\xi , \psi]=0$ which read: 
$\xi=(1,0,0,0)$, $\psi=(0,0,0,1)$. Since 
$g_{\mu \nu}=g_{\mu \nu}(r,\theta)$, there are two quantities 
that are conserved along the geodesics

\begin{eqnarray}
p_t&=& \frac{\partial \mathcal{L}}{\partial \dot{t}} = 
g_{tt} \dot{t} + g_{t \phi} \dot{\phi}= 
g_{tt} U^t + g_{t \phi} U^{\phi}= -E, \nonumber \\
p_{\phi} &=& \frac{\partial \mathcal{L}}{\partial \dot{\phi}} = 
g_{t \phi} \dot{t} + g_{\phi \phi} \dot{\phi}= 
g_{t \phi} U^t + g_{\phi \phi} U^{\phi}= L,
\label{constants}
\end{eqnarray}

\noindent 
where $U^{\mu}=(U^t,U^r,U^{\theta},U^{\phi})$ is the 4-velocity which is 
normalized to unity rendering

\begin{eqnarray}
-1 &=& g_{tt} (U^t)^2 + g_{rr} (U^r)^2 + g_{\theta \theta} (U^{\theta})^2+
g_{\phi \phi} (U^{\phi})^2 \nonumber \\
&& + g_{t \phi} U^t U^{\phi}.
\label{condition}
\end{eqnarray}

\noindent 
Two of these 4-velocity components can be found by inverting 
(\ref{constants})

\begin{equation}
U^t = \frac{g_{\phi \phi} E + g_{t \phi} L}{g_{t\phi}^2 -g_{tt} g_{\phi \phi}}
\quad , \quad
U^{\phi} = -\frac{g_{t \phi} E + g_{tt } L}{g_{t\phi}^2 -g_{tt} g_{\phi \phi}}.
\label{us}
\end{equation}

\noindent 
Inserting (\ref{us}) in (\ref{condition}) one obtains

\begin{equation}
g_{rr} \left ( U^r \right )^2 + V_{eff} =0,
\label{veff}
\end{equation}

\noindent 
where $V_{eff}$ is an effective potential given by

\begin{equation}
V_{eff}=
1+ g_{\theta \theta} \left ( U^{\theta} \right )^2 -
\frac{E^2 g_{\phi \phi}+ L^2 g_{tt} + 2 E L g_{t \phi}}
{g_{t \phi}^2-g_{tt}g_{\phi \phi}}.
\label{veffexplicit}
\end{equation}

\noindent 
The goal is to write the parameters of an axialsymmetric 
space-time in terms of 
the observational red and blue shifts $z_{red}$ and $z_{blue}$ of light
emitted by massive particles moving around a compact object.
These photons have 4-momentum $k^{\mu}=(k^t,k^r,k^{\theta},k^{\phi})$ 
that move along null geodesics $k_{\mu} k^{\mu}=0$. Using the same Lagrangian
(\ref{lagrangian}) one gets two conserved quantities

\begin{eqnarray}
-E_{\gamma} &=& g_{tt} k^t + g_{t\phi} k^{\phi}, \nonumber \\
L_{\gamma} &=& g_{\phi t} k^t + g_{\phi \phi} k^{\phi}.
\label{null}
\end{eqnarray}

The frequency shift $z$ associated to the emission and detection of 
photons is defined as

\begin{equation}
1+z= \frac{\omega_e}{\omega_d},
\label{zzz}
\end{equation}

\noindent where $\omega_e$ is the frequency emitted by an observer moving with 
the massive particle at the emission point $e$ and $\omega_d$
the frequecy detected by an observer far away from the source
of emission. These frequencies are given by

\begin{equation}
\omega_e = - k_{\mu} U^{\mu} |_e \quad , \quad
\omega_d = - k_{\mu} U^{\mu} |_d.
\label{emission}
\end{equation}

\noindent 
$U^{\mu}_e$ and $U^{\mu}_d$ are the 4-velocity of the emisor and
detector respectively. If the detector is located very far away from
the source ($r \to \infty$) then 
$U^{\mu}_d=(1,0,0,0)$ since $U^r_d,U^{\theta}_d,U^{\phi}_d \to 0$,
whereas $U^t=E=1$. The frequency $\omega_e= -k_{\mu}U^{\mu}|_e$ is 
explicitly given by\\

$\omega_e = \left ( E_{\gamma}U^t-L_{\gamma}U^{\phi}-g_{rr} U^r k^r-
g_{\theta \theta} U^{\theta} k^{\theta} \right)|_e $,
\\

\noindent 
with a similar expression for $\omega_d$. As a result (\ref{zzz})
becomes

\begin{equation}
1+z=
\frac{\left ( E_{\gamma}U^t-L_{\gamma}U^{\phi}-g_{rr} U^r k^r-
g_{\theta \theta} U^{\theta} k^{\theta} \right)|_e}
{ \left ( E_{\gamma}U^t-L_{\gamma}U^{\phi}-g_{rr} U^r k^r-
g_{\theta \theta} U^{\theta} k^{\theta} \right)|_d }.
\label{oneplusz}
\end{equation}

\noindent This is an expression for the red and/or blue shifts of light
emitted by massive particles that are orbiting around a compact object
measured by a distant observer. The apparent impact parameter
$b \equiv \frac{L_{\gamma}}{E_{\gamma}}$ of photons, that is to say,
the minimum distance to the origin $r=0$ was introduced for
convenience. Due to the fact that
$E_{\gamma}$ and $L_{\gamma}$ are preserved along null geodesics all
the way from emission to detection one has that $b_e=b_d$. On the
other hand, a set of massive particles (that could be a set of stars)
that may be orbiting around a compact object (that could be a black
hole) is expanding as a whole and it has a redshift
$z_c$. Yet all those particles are individually moving having
therefore, an individual redshift. Astronomers define a kinematic 
redshift as $z_{kin}=z-z_c$, and some report their data in terms of $z_{kin}$. 
$z_c$ corresponds to a frequecy shift of a photon emitted by a 
static particle located at $b=0$ thus

\begin{equation}
1+z_c=\frac{(E_{\gamma} U^t)|_e}{(E_{\gamma} U^t)|_d}=\frac{U^t_e}{U^t_d} 
\end{equation}

\noindent
The kinematic redshift $z_{kin}=(1+z)-(1+z_c)$ can be written as

\begin{equation}
z_{kin} =
\frac{(U^t-bU^{\phi}-\frac{1}{E_{\gamma}}g_{rr}U^rk^r-\frac{1}{E_{\gamma}}
  g_{\theta \theta} U^{\theta}k^{\theta})|_e}
{(U^t-bU^{\phi}-\frac{1}{E_{\gamma}}g_{rr}U^rk^r-\frac{1}{E_{\gamma}}
  g_{\theta \theta} U^{\theta}k^{\theta})|_d}-\frac{U^t_e}{U^t_d}
\label{zcine}
\end{equation}

\noindent
The analysis can be performed with either $z_{kin}$ using (\ref{zcine}) 
or $z$ using (\ref{oneplusz}). We work with $z_{kin}$ in this paper. 
The general expression 
(\ref{zcine}) is simplified for circular orbits ($U^r=0$) in the 
equatorial plane ($U^{\theta}=0$)

\begin{equation}
z_{kin}= \frac{U^t U_d^{\phi} b_d- U^t_d U^{\phi}_e b_e}
{U^t_d(U^t_d-b_d U^{\phi}_d)}.
\label{zkin}
\end{equation}

In (\ref{zkin}) what is still needed is to take into account light 
bending due to gravitational field, that is to say, to find 
$b=b(r)$. 
The criteria employed in \cite{ulises} to construct this mapping 
is to choose the maximum value
of $z$ at a fixed distance from the observed center of the source
at a fixed $b$. Inverting (\ref{null}) to obtain
$k^{\mu}=k^{\mu}(g_{\alpha \beta},E,L)$ and inserting this expression
into $k_{\mu} k^{\mu}=0$ with $k^r=0$ and $k^{\theta}=0$  one arrives at

\begin{equation}
b_{\pm}= \frac{-g_{t \phi} \pm \sqrt{g_{t \phi}^2-g_{tt}g_{\phi \phi}}}{g_{tt}},
\label{bmm}
\end{equation}

\noindent $b_{\pm}$ can be evaluated at the emissor or detector
position. Since in general there are two different values of
$b_{\pm}$, there will be two different values of $z$ of photons
emitted by a receding ($z_1$) or an approaching object ($z_2$) with
respect to a distant observer. These kinematic shifts of photons emitted
either side of the central value $b=0$ read

\begin{equation}
z_1=\frac{U^t_e U^{\phi}_d b_{d_{-}}-U^t_d U^{\phi}_e b_{e_{-}}}
{U^t_d(U^t_d-U^{\phi}_d b_{d_{-}})},
\label{z1}
\end{equation}

\begin{equation}
z_2=\frac{U^t_e U^{\phi}_d b_{d_{+}}-U^t_d U^{\phi}_e b_{e_{+}}}
{U^t_d(U^t_d-U^{\phi}_d b_{d_{+}})}.
\label{z2}
\end{equation}

In the next section we shall apply this formalism to non-rotating
compact objects.

\section{Non-rotating space-times}

In order to apply H-N approach, it is necessary to have 
a Killing tensor $K_{\mu \nu}$ of the space-time to be analyzed, this implies 
the existance of an additional constant of motion 
$C=K_{\mu \nu} U^{\mu} U^{\nu}$. $C$ is not needed in the case of
non-rotating space-times, that is to say, for
$g_{t \phi}=0$ or when particles are orbiting just on the equatorial
plane. In the present section, we study the relationship between the
observed redshift (blueshift) of photons emitted by particles traveling
along circular and equatorial paths around 
non-rotating compact objects and the mass parameter of these objects. 
Since $g_{t \phi}$ vanishes, the apparent impact parameter becomes
$b_{\pm}= \pm \sqrt{-g_{\phi \phi}/g_{t t}}$ 
and the effective potential (\ref{veffexplicit}) acquires a rather simple form

\begin{equation}
V_{eff} = 1 + \frac{E^2}{g_{t t}}+\frac{L^2}{g_{\phi \phi}}.
\label{potentialreduce}
\end{equation}

\noindent 
For circular orbits $V_{eff}$ and its derivative
$\frac{d V_{eff}}{dr}$ vanish. 
From these two conditions one finds two general expressions 
for the constants of motion $E^2$ and $L^2$ for any non-rotating
axialsymmetric space-time

\begin{equation}
E^2=-\frac{g_{tt}^2 g_{\phi \phi}^{\prime}}
{g_{tt} g_{\phi\phi}^{\prime}-g_{tt}^{\prime} g_{\phi \phi}},
\label{energy}
\end{equation}

\begin{equation}
L^2=\frac{g_{\phi \phi}^2 g_{tt}^{\prime}}
{g_{tt} g_{\phi\phi}^{\prime} - g_{tt}^{\prime}g_{\phi \phi} },
\label{angularM}
\end{equation}

\noindent 
where primes denote derivative with respect to $r$.
In order to guarantee stability of these circular orbits, 
$V_{eff}^{\prime \prime}>0$ must hold. The general expression for 
$V_{eff}^{\prime \prime}$ is

\begin{eqnarray}
V_{eff}^{\prime \prime} &=& -E^2 \left [ 
\frac{g_{tt}^{\prime \prime}g_{tt}-2 (g_{tt}^{\prime})^2}{g_{tt}^3}
\right ]
-L^2 \left [ 
\frac{g_{\phi \phi}^{\prime \prime}g_{\phi \phi}-2 (g_{\phi
    \phi}^{\prime})^2}{g_{\phi \phi}^3} \right ] \nonumber \\
&=& \frac{g_{\phi \phi}^{\prime} g_{tt}^{\prime\prime}
 - g_{tt}^{\prime} 
g_{\phi \phi}^{\prime \prime}}{g_{tt}g_{\phi \phi}^{\prime}-g_{tt}^{\prime} g_{\phi \phi}} 
+\frac{2 g_{tt}^{\prime} g_{\phi \phi}^{\prime}}{g_{\phi \phi}g_{tt}}, 
\label{segundita}
\end{eqnarray}

\noindent where (\ref{energy}) and (\ref{angularM}) were employed in the last
step. Using the explicit form of $E$ and $L$, (\ref{energy}) and 
(\ref{angularM}), in (\ref{us}) one obtains expression for the 4-velocities
in terms of solely the metric components

\begin{equation}
U^{\phi}=\sqrt{\frac{g_{t t}^{\prime}}
{g_{tt} g_{\phi\phi}^{\prime} - g_{tt}^{\prime} g_{\phi \phi}}}, \quad
U^t= -\sqrt{\frac{-g_{\phi \phi}^{\prime}}
{g_{tt} g_{\phi\phi}^{\prime} - g_{tt}^{\prime} g_{\phi \phi} }}.
\label{velocidades}
\end{equation}

\noindent 
from which the angular velocity of particles in these circular paths 
becomes

\begin{equation}
\Omega = \sqrt{-\frac{g_{tt}^{\prime}}{g_{\phi \phi}^{\prime}}}.
\label{Omega}
\end{equation}

\noindent 
Since $b_{+}= - b_{-}$, the redshift $z_1=z_{red}$ and blueshift $z_2=z_{blue}$
are equal but with opposite sign: $z_1= -z_2$, the explicit expression is

\begin{equation}
z_1=\frac{-U^t_e U^{\phi}_d b_{d_{+}}+U^t_d U^{\phi}_e
  b_{e_{+}}}{U^t_d(U^t_d+U^{\phi}_db_{d_{+}})}.
\label{zfinal}
\end{equation}

\noindent
Furthermore, if the detector is located far away from the compact object
$r_d \to \infty$, and as we mentioned before, 
$U_d^{\mu} \to (1,0,0,0)$. Thus (\ref{zfinal}) becomes

\begin{equation}
z_1 = U^{\phi}_e b_{e_{+}}= 
\sqrt{\frac{-g_{\phi \phi}g_{tt}^{\prime}}{g_{tt} (g_{tt} g_{\phi\phi}^{\prime} - 
g_{tt}^{\prime} g_{\phi \phi} )}}.
\label{zfinalFB}
\end{equation}

\subsection{Schwarzschild Black Hole}

As our first working example of a non-rotating space-time, we consider the
Schwarzschild black hole, for which the relevant metric components are
$g_{tt}= -(1-\frac{2M}{r})$ and $g_{\phi \phi}= r^2 \sin^2{\theta}$. 
Inserting these components in (\ref{zfinalFB}) with $\theta=\pi/2$
one finds

\begin{equation}
z^2=\frac{r_cM}{(r_c-2M)(r_c-3M)},
\label{zMr}
\end{equation}

\noindent which is a relationship between the measured red-shift $z$, 
the mass parameter of a Schwarzschild black hole $M$ and the radius $r_c$ 
of a massive particle's circular orbit that emitts light and of
course, $r_c>3M$. The relationship (\ref{zMr}) is equivalent to 

\begin{equation}
M= r_c \mathcal{F}(z) \quad \text{where}
\quad \mathcal{F}_{\pm}(z)=\frac{1+5z^2\pm \sqrt{1+10z^2+z^4}}{12z^2}.
\label{Mrz}
\end{equation}

\noindent On the other hand, circular orbits are stable as long as that 
$V_{eff}^{\prime \prime}>0$, from (\ref{segundita}) 
$V_{eff}^{\prime \prime}$ reads

\begin{equation}
V_{eff}^{\prime \prime}= \frac{2M(r_c-6M)}{r_c^2(r_c-2M)(r_c-3M)},
\label{V2explicita}
\end{equation}

\noindent which is positive provided that $r_c > 6 M$; 
therefore, $\frac{r_c}{M}= \mathcal{F}^{-1} > 6$ which 
is fulfilled if and only if
$|z| < 1/\sqrt{2}$ and solely for the minus sign $\mathcal{F}_{-}(z)$. 
Hence, a measurement of the redshift $z$ of light 
emitted by a particle that follows
a circular orbit of radius $r_c$ in the equatorial plane around a
Schwarzschild black hole will have a mass parameter determined by 
$M = r_c \mathcal{F}_{-}(z)$, and $z$ must be $|z|< 1/\sqrt{2}$.
The energy, angular momentum, velocities $U^t$, $U^{\phi}$ and 
the angular velocity of 
the emitter, can be computed from (\ref{energy}),
(\ref{angularM}), (\ref{velocidades}) and (\ref{Omega}) 
and written as function
of the measurable redshift $z$ and radius $r_c$ of the circular 
photons source's orbit by using (\ref{Mrz}) 

\begin{eqnarray}
E^2 &=& \frac{(r_c -2M)^2}{r_c(r_c-3M)}= 
\frac{(1-2\mathcal{F}_{-}(z))^2}{r_c(1-3\mathcal{F}_{-}(z))}, 
\nonumber \\
L^2 &=& \frac{M r^2_c}{r_c-3M}=
\frac{r_c^2\mathcal{F}_{-}(z)}{1-3\mathcal{F}_{-}(z)},
\label{ELz}
\end{eqnarray}

\begin{eqnarray}
U^t &=& \sqrt{\frac{r_c}{r_c-3M}}= \frac{1}{\sqrt{1-3\mathcal{F}_{-}(z)}},
\nonumber \\
U^{\phi} &=& \frac{1}{r_c} \sqrt{\frac{M}{r_c-3M}} = 
\frac{1}{r_c} \sqrt{\frac{\mathcal{F}_{-}(z)}{1-3\mathcal{F}_{-}(z)}},
\label{velschw}
\end{eqnarray}

\begin{equation}
\Omega = \sqrt{\frac{M}{r_c^3}}=\sqrt{\frac{\mathcal{F}_{-}(z)}{r_c^2}}.
\label{angularvelSchw}
\end{equation}

\begin{figure}[htp]
 \centerline{ \epsfysize=6.5cm \epsfbox{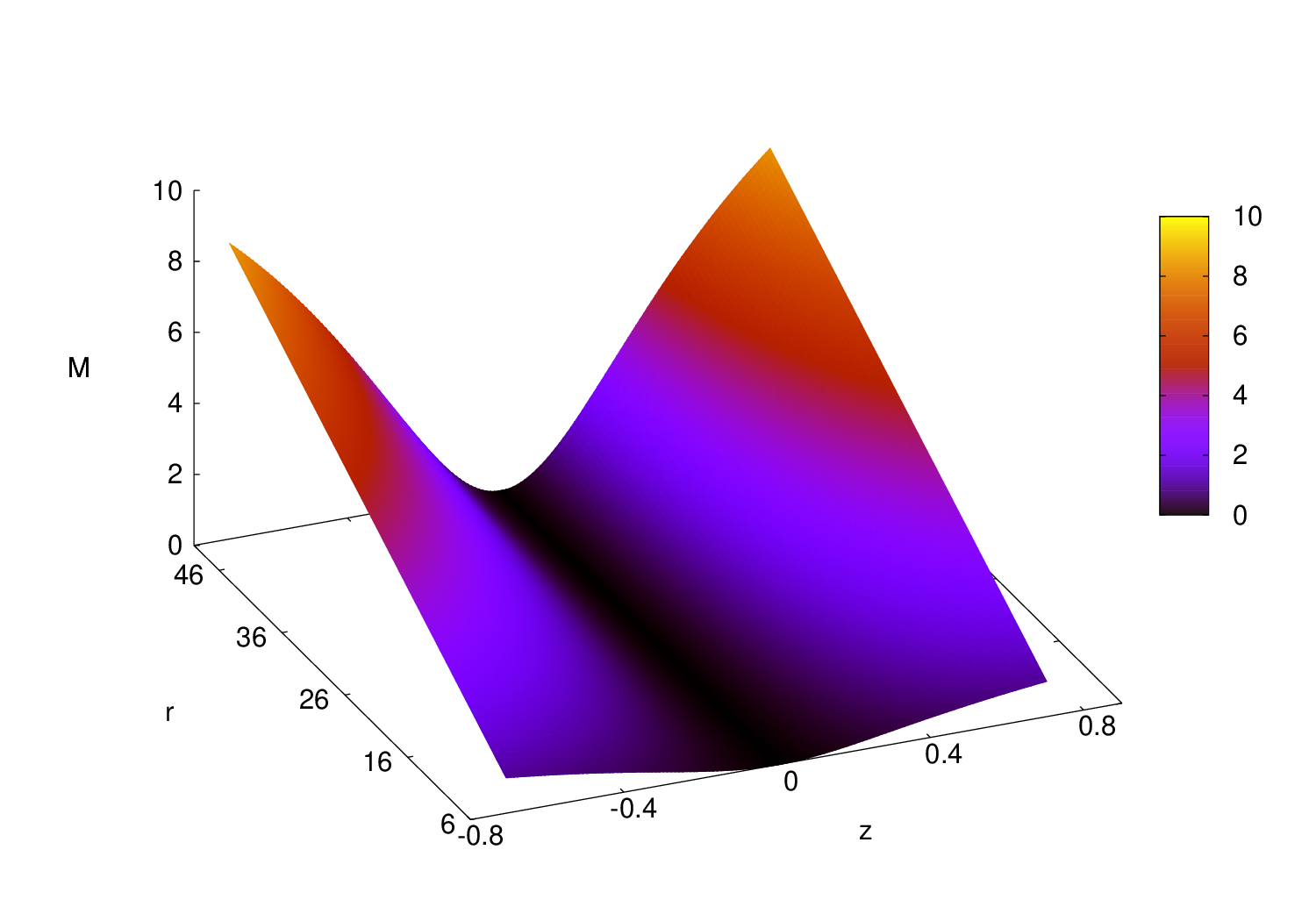}}
  \caption{It is shown the mass parameter $M$ as a function of 
    redshift ($z>0$) or blueshift ($z<0$)  
    and the radius $r$ of an eventual circular orbit of a 
    photon emitter. $M$ and $r$ are in geometrized units and scaled 
    by $pM_{\odot}$ where $p$ is an arbitrary factor of proportionality. }
  \label{MSCH}
\end{figure}

The function $M=M(r,z)=r \mathcal{F}_{-}(z)$ is in 
geometrized units (G=c=1). In order to plot it, 
we scale $M$ and $r$
by any multiple of the solar mass, this is to say, by $p M_{\odot}$,
for Srg$^*$ $p=2$.$72 \times 10^6$. 
Figure \ref{MSCH} shows this scaled relation $M=M(r,z)$ which is symmetric
with respect to the shift $z$ ($z_{red}>0$, $z_{blue}<0$).
 
Given a set of $N$ pairs $\{r,z\}_i$ of observed redshifts $z$
(blueshifts) of emitters traveling around a
Schwarzschild black hole along circular orbits of radii $r$,
a Bayesian statistical analysis might be carried out in order to
estimate the black hole mass parameter.

\subsection{The Reissner-Nordstr\"om Black Hole}

Our next non-rotating working example is the Reissner-Nordstr\"om
space-time which represents a electrically charged black hole, 
whose relevant metric components are 
$g_{tt}= -\left ( 1-\frac{2M}{r}+\frac{Q^2}{r^2}\right )$ where $Q$
is the electric charge parameter and
$g_{\phi \phi}=r^2 \sin^2{\theta}$. For circular equatorial orbits of
the photon source, the redshift reads

\begin{equation}
z^2= \frac{r_c^2(M r_c- Q^2)}{(r_c^2-3Mr_c+2Q^2)(r_c^2-2Mr_c+Q^2)}. 
\label{zRN}
\end{equation}

\noindent This relationship is equivalent to

\begin{equation}
M= r_c \mathcal{G}_{\pm}(r_c,z^2,Q^2),
\label{MasaRN}
\end{equation}

\noindent where 

\begin{eqnarray}
\mathcal{G}_{\pm} &=& \frac{1}{12z^2}\Big [ (5z^2+1)+\frac{7Q^2z^2}{r_c^2} 
\nonumber \\
 &\pm& \left ( z^4+10z^2+1 + 
\frac{z^2Q^2}{r_c^2}\left [ \frac{z^2Q^2}{r_c^2}-2(z^2+5)\right ] \right )^{1/2} 
\Big ] \nonumber \\
\label{G}
\end{eqnarray}

In this case, the conserved quantities $E^2$ and $L^2$ are

\begin{equation}
E^2= \frac{(Q^2 + r_c(r_c-2M))^2}{r_c^2(2Q^2+r_c(r_c-3M))},
\end{equation}

\begin{equation}
L^2= \frac{r_c^2(Mr_c-Q^2)}{2Q^2+r_c(r_c-3M)}.
\end{equation}

\noindent 
$E^2$ and $L^2$ are real only if 
$r_c^2-3Mr_c+2Q^2>0$ and $Mr_c-Q^2>0$. 
Therefore, $z^2$ is positive provided that $r_c^2-2Mr_c+Q^2>0$.
As it is known, in this metric, one distinguishes three regions:
$0<r<r_{-}$, $r_{-}<r<r_{+}$ and $r_{+}<r$, where 
$r_{\pm}=M \pm \sqrt{M^2-Q^2}$ are the roots of $r^2-2Mr+Q^2=0$, which are
real and distinct only if $M^2 > Q^2$ stands. The surface $r=r_{+}$
is an event horizon similar to that $r=2M$ for the Schwarzschild's
metric \cite{chandra}. Since $r>r_{+}$ implies $r^2-2Mr+Q^2>0$, 
our analysis is performed for $r>r_{+}$, that is, outside the event horizon.

\begin{figure}[htp]
 \centerline{ \epsfysize=6.0cm \epsfbox{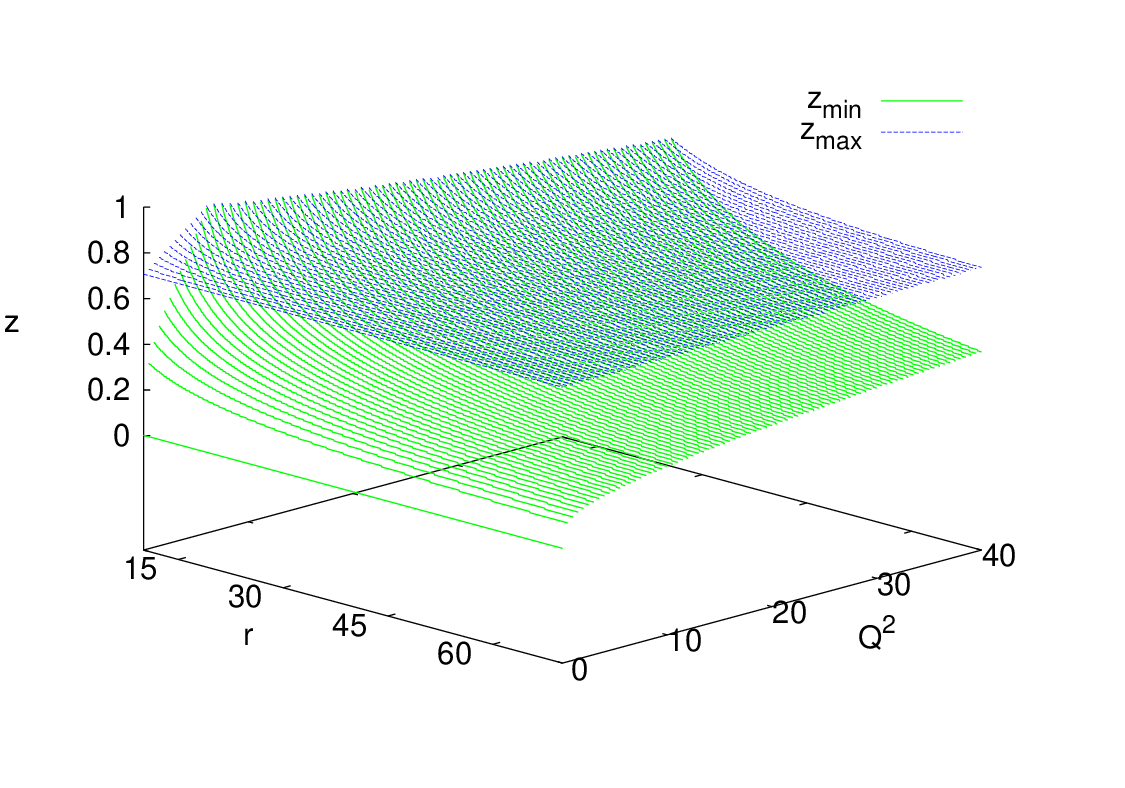}}
  \caption{Minimum $z_{min}$ and maximim $z_{max}$
    redshift surfaces as a function of the radius $r$ 
    of circular orbits followed
    by photon emitters around a Reissner Nordstr\"om black hole and 
    its charge parameter $Q^2$. Only for redshifts bounded by these surfaces,
    the corresponding values $M=M(z^2,Q^2,r)=r \mathcal{G}_{-}$ are acceptable. 
    $M$,$Q$ and $r$ are in geometrized units and scaled 
    by $pM_{\odot}$ where $p$ is an arbitrary factor of proportionality. }
  \label{qrz}
\end{figure}

The stability of circular equatorial orbits requirement 

\begin{equation}
V^{''}_{eff}= \frac{Mr_c(18Q^2+2r_c^2-12Mr_c)-8Q^4}{r_c^2(2Q^2+r_c(r_c-3M))(Q^2+r_c(r_c-2M))}
>0,
\label{VppRN}
\end{equation}

\noindent tells us that $Mr_c(9Q^2+r_c^2-6Mr_c)-4Q^4>0$. Inserting 
$M=r_c \mathcal{G}_{\pm}$ into this last condition would yield,
in principle, an inequality that may bound the values of the redshift 
$z$, as it was the case for Schwarzschild. 
This inequality turns out to be
cumbersome to be analysed analytically; hence, the analysis 
was performed numerically in the following
manner: given values of $Q^2$ and $r_c$, we vary $z^2$
and compute $M=r_c \mathcal{G}_{\pm}(z^2,Q^2,r_c)$ for each value of $z^2$.
With this value $M$ at hand, we check whether  
the four conditions are all satisfied:
(i) $M^2 > Q^2$, (ii) $r^2 - 3Mr+2Q^2>0$, (iii) $Mr-Q^2>0$ and 
(iv) $Mr(9Q^2+r^2-6Mr)-4Q^4>0$. The second and third 
inequalities guarantee that, 
one indeed, has circular and equatorial orbits, the fourth stems
from $V_{eff}^{\prime \prime}>0$. We look for the minimum and maximum value
of $z$ for which these four condictions are fulfilled. 
This process is repeated for several values of $Q^2$ and $r_c$. 
For $Q=0$, the result for Schwarzschild ($|z|<1/\sqrt{2}$) is
recovered. Figure \ref{qrz} shows the surfaces $z_{min}=z_{min}(r_c,Q^2)$ and
$z_{max}=z_{max}(r_c,Q^2)$. Only for frequency shifts $z$ such that
$|z| \in (z_{min},z_{max})$, the corresponding values 
$M=M(z^2,Q^2,r_c)=r_c \mathcal{G}_{-}$ are acceptable. 

The velocities $U^{\phi}$ and $U^t$ of photons emitters orbiting in 
circular and equatorial paths are

\begin{eqnarray}
U^{\phi} &=&  \frac{1}{r_c} \sqrt{\frac{Mr_c-Q^2}{r_c^2(2Q^2+r_c(r_c-3M))}},
\nonumber \\
U^t &=& \sqrt{\frac{r_c^2}{2Q^2+r_c(r_c-3M)}},
\end{eqnarray}

\noindent and their angular velocity is given by

\begin{equation}
\Omega= \sqrt{\frac{Mr_c-Q^2}{r_c^4}}.
\end{equation}

\noindent Since $M=r_c\mathcal{G}_{-}(z^2,r_c,Q^2)$, these 4-velocity 
components and $\Omega$ are actually functions of the redshift $z$, 
the radius of the circular orbit $r_c$ and the parameter $Q^2$. 
Unlike the Schwarzshild black hole, there is not an analytic 
relationship of the mass parameter $M$ in terms only of 
the measurable variables $z$ and $r$, it depends also on $Q^2$.
At any rate, given a set of the observables $\{ z,r \}_i$,
Bayesian statistical analysis would provide an estimate for both parameters
$M$ and $Q$.

\subsection{Boson Stars}

Colpi  {\it et al} performed a study of self-interacting Boson
stars which were modeled by a complex scalar field endowed
with a quartic potential 
$V=\frac{m^2}{2}|\phi|^2 + \frac{\lambda}{4}|\phi|^4$. The stability
analysis yielded equilibrium configurations along either an stable and
unstable branch \cite{Colpi,Ruffini}. We will be concerned with stable
equilibrium configurations of Boson stars for which
the metric reads

\begin{equation}
ds^2= -\alpha^2(r) dt^2+a^2(r) dr^2 + 
r^2 \left ( d\theta^2+ \sin^2{\theta} d \varphi^2 \right ).
\label{fbs}
\end{equation}

\noindent The components $g_{rr}=a^2(r)$ and $g_{tt}= -\alpha^2(r)$ 
are found by solving 

\begin{eqnarray}
\frac{da}{dx} =\frac{a}{2}\left[\frac{1-a^2}{x}+
  a^2x\left(\left[\frac{\Omega^2}{\alpha^2}+1+\frac{\Lambda}{2}\hat{\phi}^2\right]
 \hat{\phi}^2 + \frac{\hat{\phi}^{'2}}{a^2}\right)\right], \nonumber \\
\frac{d\alpha}{dx}=\frac{\alpha}{2}\left[\frac{a^2-1}{x}+
  a^2x\left(\left[\frac{\Omega^2}{\alpha^2}-1-\frac{\Lambda}{2} \hat{\phi}^2\right]
   \hat{\phi}^2 + \frac{\hat{\phi}^{'2}}{a^2}\right)\right], \nonumber \\
\label{CEquilibrium}
\end{eqnarray}

\noindent
where, for numerical purposes, we have introduced the following
dimensionless variables: $x=mr$, 
$\hat{\phi}=\sqrt{4\pi G}\phi$, $\Lambda= \lambda/4\pi Gm^2$ and 
$\Omega= \omega/m$, where $m$ is 
the mass of complex scalar field $\phi$, $\omega$ its 
frequency and $\lambda$ the dimensionless self-coupling of the
scalar. Here $'$ represents the derivative with respect to $x$.

For the complex scalar field, we consider a harmonic form
$\Phi(t,r)=\phi(r)e^{-i\omega t}$ and solve the Klein-Gordon
equation, that in terms of the dimensionless variables, takes the form

\begin{equation}
\hat{\phi}^{''}= \left(1 -
\frac{\Omega^2}{\alpha^2}+\Lambda\hat{\phi}^2\right)a^2\hat{\phi} -
\left(\frac{\alpha^{'}}{\alpha}-\frac{a^{'}}{a}+\frac{2}{x}\right)
\hat{\phi}^{'}.
\end{equation}

The boundary conditions for the metric functions and the scalar field,
in order to guarantee regularity at the origin and asymptotic flatness
at infinity, are: $a(0)=1$, $\alpha(0)=1$, $\phi(0)=\phi_0$,
$\phi^{'}(0)=0$, 
$\displaystyle{\lim_{x\rightarrow \infty} }\alpha(r) = \displaystyle{\lim_{x
  \rightarrow \infty}}1/a(x)$ and $\displaystyle{\lim_{x\rightarrow \infty}} \phi(x)
\approx 0$. 

\begin{figure}[htp]
  \centerline{ \epsfysize=6.0cm \epsfbox{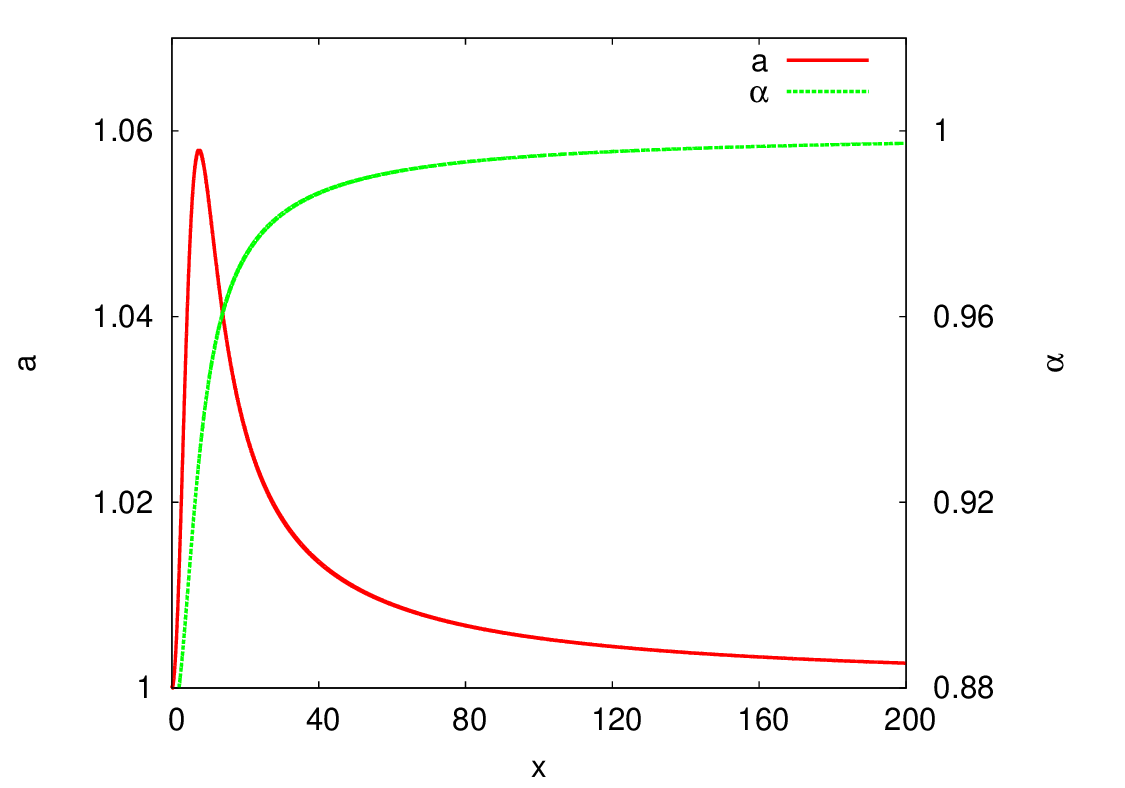}}
  \caption{Metric functions $g_{tt}= -\alpha^2(x)$ and  
    $g_{rr}=a^2(x)$, of equilibrium configurations for Boson stars,
    corresponding to the values of the quartic parameter
   $\Lambda=0$.}
  \label{metric}
\end{figure}

The system is basically an eigenvalue problem
for the frequency of  the boson star $\omega$ as a function of a
parameter, the so called, central value of the scalar field
$\phi_0$ which determines the mass $M$ of a boson star. 
This system can be solved by using the shooting method \cite{NR}.
Figure \ref{metric} shows the metric component $g_{tt}= -\alpha^2(x)$ and 
$g_{rr} = a^2(x)$ for Boson stars with $\Lambda=0$.

For circular orbits ($\dot{x}=0$) with radius $x_c$
the effective potential and its derivative vanish.
From these conditions $L^2$ and $E^2$ are obtained 

\begin{equation}
L^2=\frac{x_c^3 \alpha^{\prime}(x_c)}{\alpha(x_c) - x_c \alpha^{\prime}(x_c)}
, \quad
E^2=\frac{\alpha^3(x_c)}{\alpha(x_c) - x_c \alpha^{\prime}(x_c)},
\label{FBL2E2}
\end{equation}

\noindent
here $x_c=mr_c$. Choosing $E^2$ and $L^2$ as in 
(\ref{FBL2E2}) guarantees circular orbits. 
Generally both, $\alpha$ and $\alpha^{\prime}$ are non-negative; therefore, 
given a numerical solution, we only need to determine the domain
$\mathcal{D}$ of the radial variable $x$ 
where $\alpha - x \alpha^{\prime} > 0 $ and work exclusively in that domain.
We then compute the values $L^2$ and $E^2$ with (\ref{FBL2E2}) and 
perform a survey in $\mathcal{D}$ checking where the condition
for stable circular orbits $V_{eff}^{\prime \prime}>0$ holds.
Thereby one finds a set of parameters $\{ (E,L,x_c) \}$ which give us 
circular orbits, $x_c \epsilon \mathcal{D}$. 

According to the equation (\ref{zfinalFB}), the redshift of photons
emitted by particles orbiting a boson star is calculated by

\begin{equation}
z(x) = \sqrt{\frac{x\alpha^{'}}{\alpha^2(\alpha - x\alpha^{' })}}.
\end{equation}

\begin{figure}[ht]
  \begin{center}
      \resizebox{80mm}{!}{\includegraphics{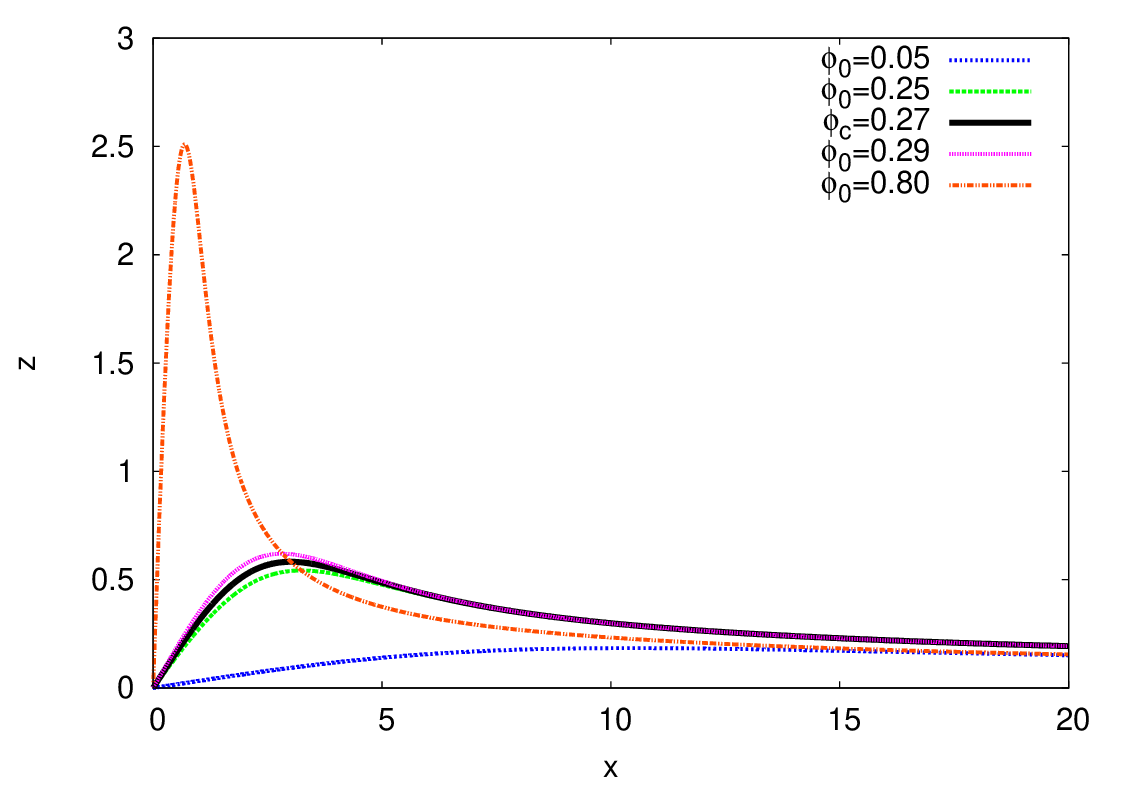}}
      \\ 
      \resizebox{80mm}{!}{\includegraphics{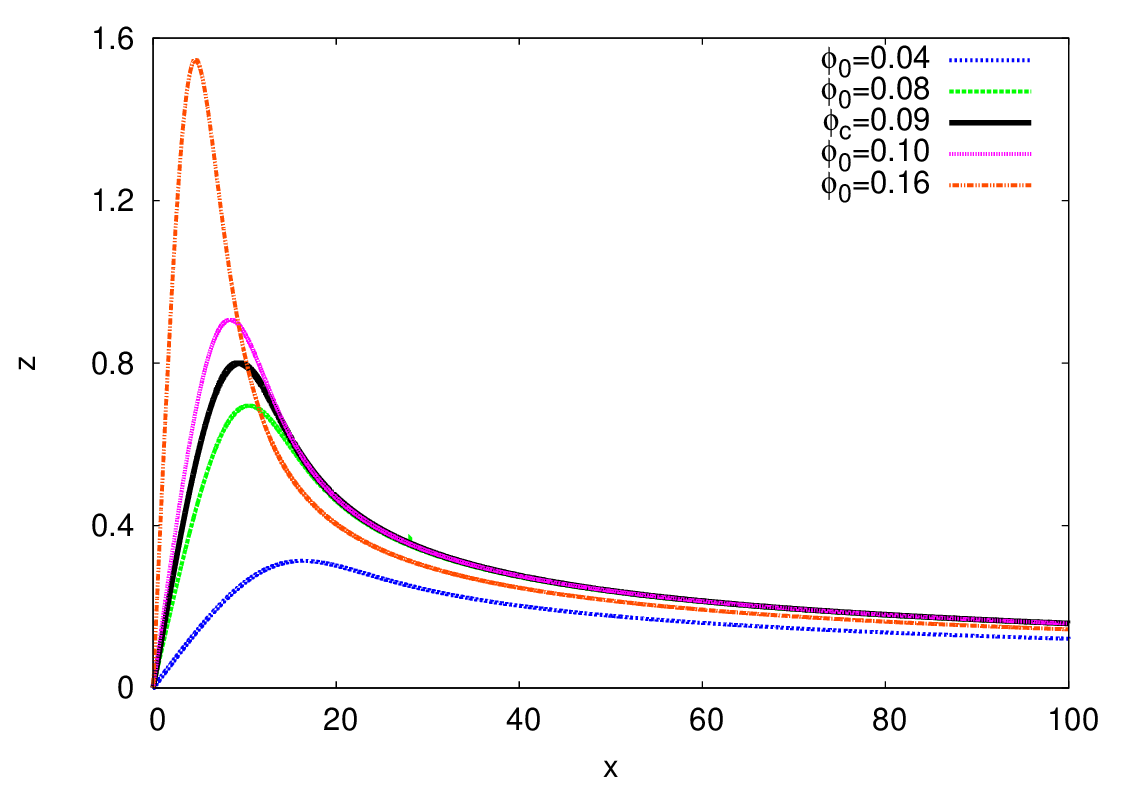}}
      \caption{We show the redshift of photons emitted by particles
        orbiting boson stars with different masses corresponding to
        different central values $\phi_0$,
        as function of the scaled radius of the orbit. $\phi_c$
        is the central value for the maximun mass. The upper plot 
        corresponds to the case $\Lambda=0$ and the lower plot to
        $\Lambda=100$.}
      \label{shiftbosones}
  \end{center}
\end{figure}

Figures \ref{shiftbosones} show the $z$ behavior in function of
$x$ for several boson stars with different masses and for two
values of $\Lambda$, $0$ and $100$. The solid black
curve represents the boson star correspoding to the critical mass
$M_{crit}$. For $M<M_{crit}$, or equivalently, $\phi_0 < \phi_{crit}$
the boson star is stable, otherwise is unstable. $M_{crit}=0$.$633$ and
$M_{crit}=2$.$254$ for $\Lambda= 0$ and $\Lambda =100$
respectively. 
In figure \ref{shiftbosones} for $\Lambda=0$, it is observed that the
maximum value of the redshift increases as
the central value $\phi_0$ of the scalar field increases.  
But for large values of $x$, all the curves $z(x)$ seem to get closer to the 
value of $z_{crit}(x)$ at large $x$ for a boson star with the critical 
mass $M_{crit}$. One also can observe that the curves $z(x)$
corresponding to smaller masses than the critical, remain below
the solid black curve $z_{crit}(x)$. 

The table below, shows the values of the masses
corresponding to stable and unstable boson stars 
for both $\Lambda=0$ and $\Lambda=100$.\\

\begin{tabular}{|c|c|c|c||c|c|c|c|}
\hline
\multicolumn{4}{|c||}{$\Lambda=0$} & \multicolumn{4}{|c|}{$\Lambda=100$}\\
\hline
\multicolumn{2}{|c|}{Stable } &\multicolumn{2}{|c||}{Unstable} &\multicolumn{2}{|c|}{Stable } &\multicolumn{2}{|c|}{Unstable}\\
\hline
$\phi_0$ & $M_T$ & $\phi_0$ & $M_T$ & $\phi_0$ & $M_T$ & $\phi_0$ & $M_T$\\
\hline
0.05 & 0.416 & 0.29 & 0.620 & 0.04 & 1.371 & 0.10 & 2.249\\
\hline
0.25 & 0.620 & 0.80 & 0.431  & 0.08 & 2.227 & 0.16 & 1.892\\
\hline
\end{tabular} \\

\begin{figure}[htp]
  \centerline{ \epsfysize=6cm \epsfbox{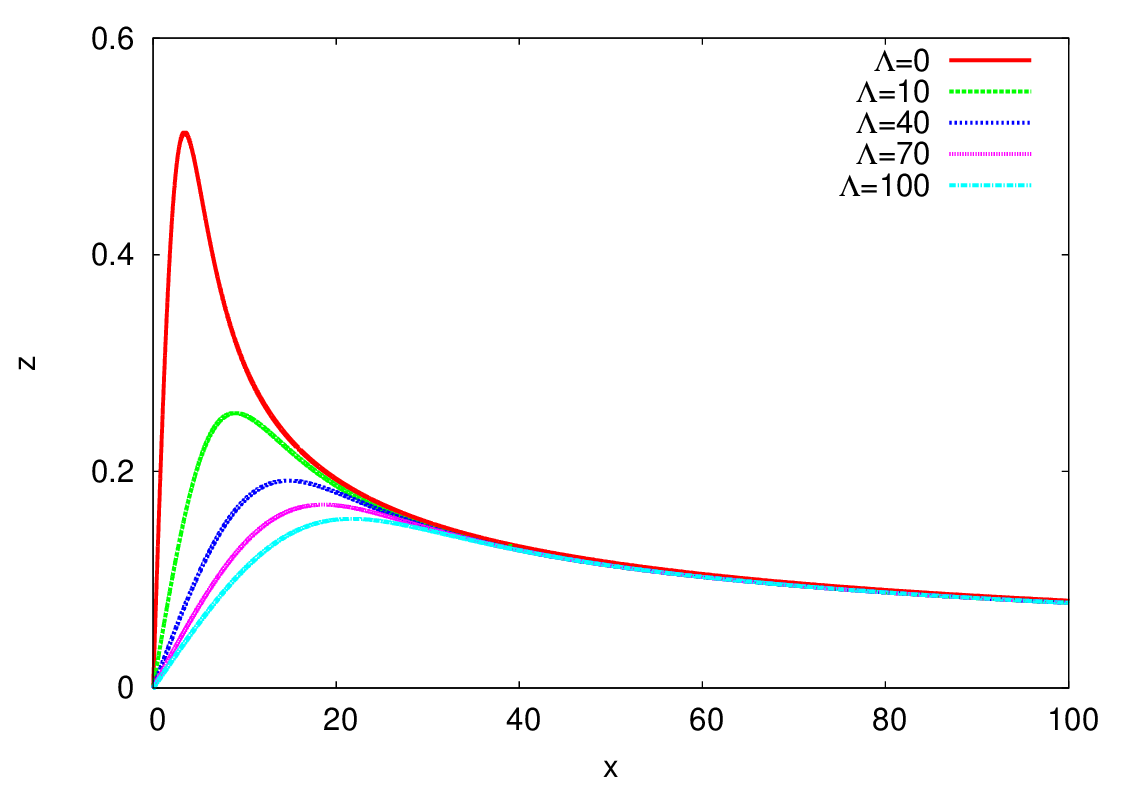}}
  \caption{Redshift due to a particle orbiting  different 
    self-interacting boson stars with the same mass $M_T=0.63$.}
  \label{shiftlambda}
\end{figure}

\noindent One can also note that for configurations 
with the same value of mass, but
different self-interacting parameter, the maximun redshift increases
as $\Lambda$ decreases. For large values of $x$, the redshift for
all configurations converge to the same values (see
fig. \ref{shiftlambda}).

\section{Kerr Black Hole}

Explicit expressions for the shifts $z_1$ and $z_2$ computed at 
either side of $b=0$ were found by H-N

\begin{eqnarray}
z_1&=&\frac{\pm \sqrt{M}\left (2aM + r_c\sqrt{r_c^2-2Mr_c+a^2}\right ) }
{r_c^{3/4}(r_c-2M)\sqrt{r_c^{3/2}-3Mr_c^{1/2}\pm 2a M^{1/2}} }, 
\nonumber \\
z_2&=& \frac{\pm \sqrt{M}\left (2aM - r_c\sqrt{r_c^2-2Mr_c+a^2}\right ) }
{r_c^{3/4}(r_c-2M)\sqrt{r_c^{3/2}-3Mr_c^{1/2}\pm 2a M^{1/2}} }. 
\label{zskerrEqn}
\end{eqnarray}

\noindent Upper signs corresponds 
to co-rotating orbits and lower signs to counter-rotating 
orbits. From (\ref{zskerrEqn}) the rotating parameter $a$ as a function
of the mass parameter $M$, the radius of circular equatorial orbits $r_c$
of particles around the Kerr black hole emitting light and the corresponding 
$z_1$ and $z_2$ turns out to be

\begin{equation}
a^2(\alpha,\beta,r_c,M) = \frac{r_c^3(r_c-2M)\alpha}{4M^2 \beta-r^2_c \alpha}, 
\label{a2}
\end{equation}

\noindent
where $\alpha \equiv (z_1 + z_2)^2$ and $\beta \equiv (z_1 - z_2)^2$. 
Nonetheless, there is not an 
explicit expression to find the mass parameter $M$, instead, 
there is an eight order polynomial for it derived also from (\ref{zskerrEqn}). 
In this section, we 
carry out a numerical analysis to study how $M$ varies
with $r_c$, and the shifts $z_1$ and $z_2$ detected by a far away observer.
The metric components of the Kerr black hole in the Boyer-Lindquist
coordinates are given by

\begin{eqnarray}
g_{tt}&=&-\left ( 1- \frac{2Mr}{\Sigma} \right ), \quad
g_{t \phi}= -\frac{2Mar\sin^2{\theta}}{\Sigma}, \nonumber \\
g_{\phi \phi} &=& \left (r^2+a^2+\frac{2Ma^2r \sin^2{\theta}}{\Sigma}
\right ) \sin^2{\theta}, \nonumber \\
g_{rr} &=& \frac{\Sigma}{\Delta}, \quad g_{\theta \theta}= \Sigma,
\label{Kerr}
\end{eqnarray}

\noindent
where 

\begin{equation}
\Delta \equiv r^2 + a^2 - 2Mr, \quad \Sigma \equiv r^2 + a^2
\cos^2{\theta},
\nonumber
\end{equation}

\noindent with the restriction $M^2 \ge a^2$. For circular and 
equatorial orbits, the two conserved quantities are  \cite{Bardeen}

\begin{eqnarray}
E &=& \frac{r^{3/2}-2M\sqrt{r}\pm a\sqrt{M}}
{r^{3/4}\sqrt{r^{3/2}-3M\sqrt{r}\pm 2a \sqrt{M}} }, \nonumber \\
L &=& \frac{\pm \sqrt{M}(r^2\mp 2a\sqrt{Mr} +a^2)}
{r^{3/4}\sqrt{r^{3/2}-3M\sqrt{r}\pm 2a \sqrt{M}} }. 
\label{EL}
\end{eqnarray}

Co-rotating orbits (upper signs) have $L>0$ whereas 
counter-rotatings (lower signs) orbits have $L<0$.
In order to have real values for $E$ and $L$, and thereby
circular orbits, it is necessary that 

\begin{equation}
r^{3/2}-3M\sqrt{r}\pm 2a \sqrt{M} \ge 0.
\label{LEreales}
\end{equation}

Circular-equatorial orbits can be either bound or unbound. The
later type are those for which, given a small outward perturbation,
the particle will go to infinity, one has bound orbits
otherwise. There are bound orbits provided that

\begin{equation}
r > r_{mb}= 2M \mp a + 2 \sqrt{M} \sqrt{M\mp a}
\label{rmb}
\end{equation}

\noindent is satisfied. Not all bound orbits are stable,
only those whose radius satisfies $V_{eff}^{\prime \prime}(r) \ge 0$ 
are stable \cite{Bardeen}. This condition is akin to

\begin{eqnarray}
&& r \ge r_{ms}= M \left [ 3 +Z_2 \mp \sqrt{(3-Z_1)(3+Z_1+2Z_2)} \right ],
\nonumber \\
&& Z_1 \equiv 1+\left ( 1-\frac{a^2}{M^2} \right )^{1/3}
\left [ \left (1+\frac{a}{M} \right )^{1/3}
+ \left (1-\frac{a}{M} \right )^{1/3} \right ],
\nonumber \\
&& Z_2 \equiv \sqrt{ 3 \frac{a^2}{M^2} + Z_1^2}.
\label{rms}
\end{eqnarray}

$M$ can not be written as an explicit function of $r_c$, $\alpha$ and $\beta$,
or equivalently as a function of $r_c$, $z_1$ and $z_2$.
In order to find the mass parameter $M$, one has to numerically find the 
roots of the eight order polynomial derived from (\ref{zskerrEqn})

\begin{eqnarray}
&& F(M)=
\left [ 16 r_c M^3 -(4 \beta M^2-\alpha r_c^2)(r_c-2M)(r_c-3M) 
\right ]^2 \nonumber \\
&& - 4 \alpha r_c^2 M(r_c-2M)^3(4\beta M^2-\alpha r_c^2).
\label{PolM}
\end{eqnarray}

It is convenient to normalized $M$ by an arbitrary $M_{max}$ as
$\tilde{M}= M/M_{max}$ thereby $0<\tilde{M}\le 1$. $r_c$ is also scaled
with $M_{max}$ as $\tilde{r}_c = r_c/M_{max}$. 
$M_{max}$ may be chosen again as $p M_{\odot}$.
We will work with the $\tilde{M}$ and $\tilde{r}$ 
variables henceforth but we will drop the tildes.

For a given value of the
radius of the emitter's circular path $r_c$, one sets the size of the
parameter domain 
$\mathcal{D}=(z_{1min},z_{1max}) \times (z_{2min},z_{2max})$ 
where a search of these polynomial's roots is carried out. 
The polynomial (\ref{PolM}) has the following properties:
$F(M;r_c,z_1,z_2)= F(M;r_c,z_2,z_1)= F(M;r_c,-z_1,-z_2)$ which
is useful for choosing $\mathcal{D}$. Recalling that 
the two different values of $z$ correspond to photons
emitted by a receding ($z_1$) or an approaching object ($z_2$) with
respect to a distant observer, an apposite domain would be
$\mathcal{D}= (0,z_{1max}) \times (-z_{2min},0)$.
At each point $q=(z_1,z_2) \in \mathcal{D}$, (\ref{PolM}) is
numerically solved to attain $M=M(q;r_c)$.
One starts with a given fixed value of $r_c$, and search in our domain 
$\mathcal{D}$ for the subset 
$\mathcal{D}_{r_c}$  where roots of $F(M;r_c,q)=0$ exist.
In principle, there may be up to eight real roots $M_i$ (or none) at 
$q \in \mathcal{D}$. If there is at least one root, the corresponding
$a^2$ is computed using (\ref{a2}) and we test whether $M^2 \ge a^2$
actually holds. If this is the case,
$r^{3/2}-3M\sqrt{r}\pm 2a \sqrt{M} \ge 0$ should be tested to determine
for which roots of $P(M;r_c,q)$ there is indeed, a circular 
orbit. Moreover, this inequality
tells us what type of orbit we are dealing with at $q$, 
either a co or counter-rotating one. 
We discard those roots of the polynomial (\ref{PolM}) at a point 
$q \in \mathcal{D}$ that do not fulfill the conditions for circular,
bound ($r>r_{mb}$) and stable ($r>r_{rms}$) orbits.
What we have found is that, not in every single point 
$q \in \mathcal{D}$, there is a root of $F(M)=0$ that leads us 
to a circular stable orbit of radius $r_c$ followed by a photon
emitter particle, only in a subset $\mathcal{D}_{r_c} \subset \mathcal{D}$ 
such a mass parameter exists. 

\begin{figure}[htp]
  \centerline{ \epsfysize=6cm \epsfbox{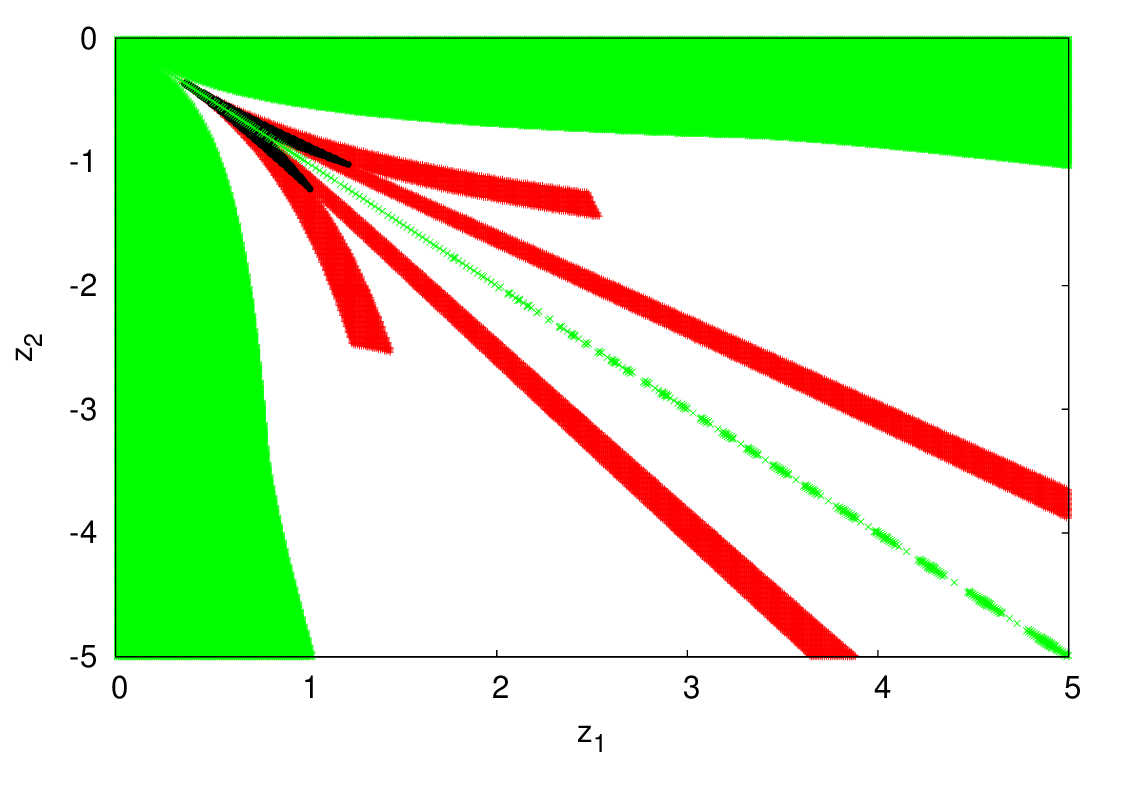}}
  \caption{For corotating orbits around the Kerr black hole, at each 
    point in the red region of the red-blueshift space $(z_1,z_2)$,
    there is a single root ($M$) of the polynomial 
    $F(M;r_c,z_1,z_2)=0$ for $r_c=3$.
    At each point of the small black region there are two roots. In the white 
    region, there exist a mass parameter $M$, yet it does not correspond 
    to a stable orbit. In the green region, there is no root at all.}
  \label{zskerr}
\end{figure}

\begin{figure}[htp]
  \centerline{ \epsfysize=7.5cm \epsfbox{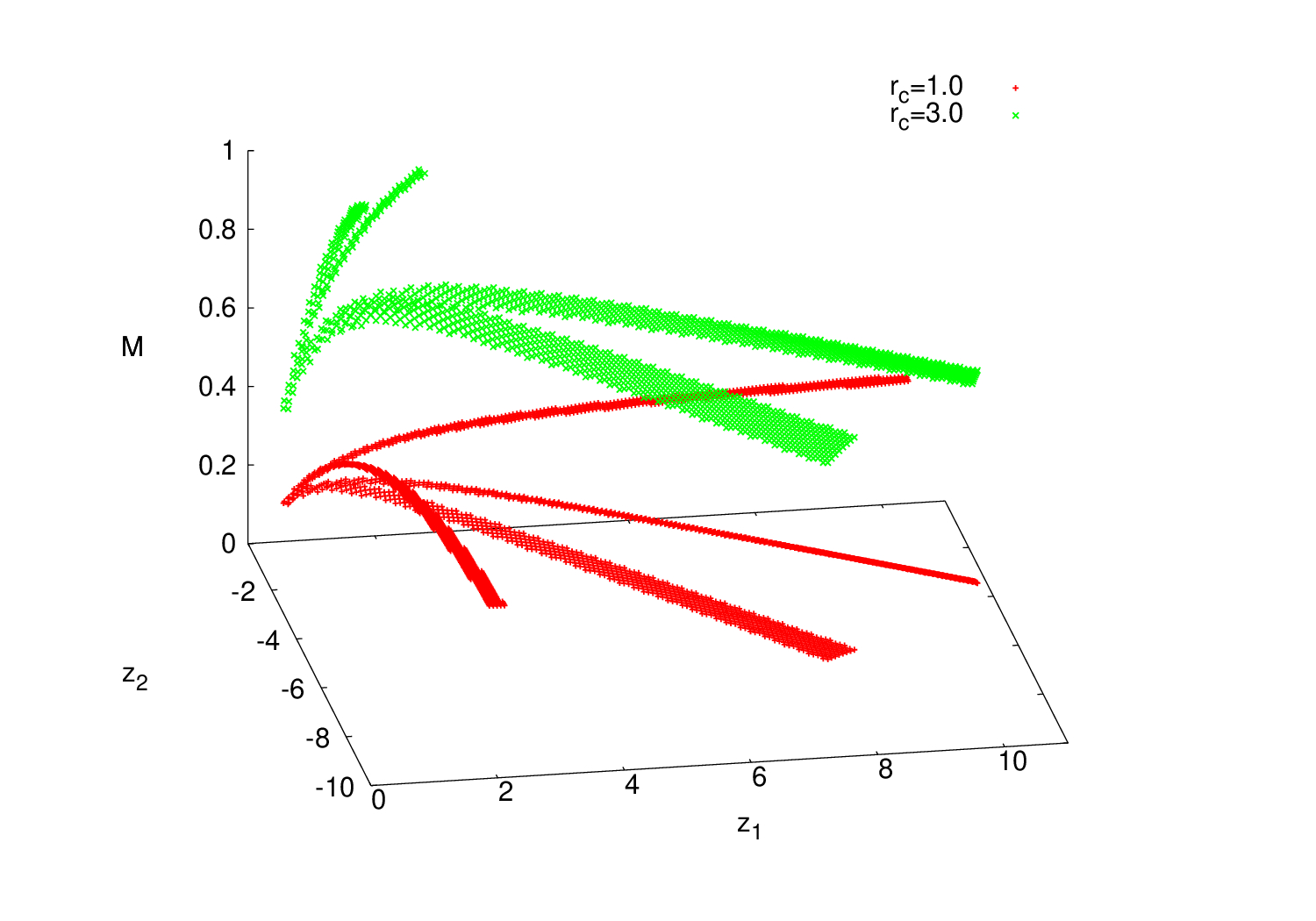}}
  \caption{Mass parameter values of a Kerr black hole 
    obtained by solving the polynomial $F(M)$, equation (\ref{zskerrEqn}),
    for a set of points $\{ (z_1 , z_2, r_c ) \}$, for two values of $r_c$. 
    As $r_c$ increases, the domain $\mathcal{D}_{r_c}$ where roots of 
    $F(M)$ exist for stable circular orbits shrinks. $M$ and $r_c$ 
    are scaled by $p M_{\odot}$.
    }
  \label{Masa05}
\end{figure}

Furthermore, in all the surveys we have done 
on domains with different sizes and different values of $r_c$, 
in almost every point 
$q \in \mathcal{D}_{r_c}$, the mass $M$
obtained is unique, so is the rotation parameter $a$. There is a 
tiny region $\mathcal{D}_{double} \subset \mathcal{D}_{r_c}$ 
where two roots at $q \in \mathcal{D}_{r_c}$ exist, these two roots are
very close to each other, the difference between each pair, 
is typicaly of order $10^{-2}$ or smaller. Figure \ref{zskerr} 
shows the bounds of the frequency shifts
where there is a mass parameter corresponding to circular stable 
corotating orbits of photon emitters. In the subset 
$\mathcal{D}_{r_{c}}$ of the parameter space $(z_1,z_2)$  
there is a single (red region) and a double (black region)
root ($M$) of the polynomial $F(M;r_c,z_1,z_2)=0$ for $r_c=3$. 
There is a rather small region in $\mathcal{D}$ where retrograde 
orbits are allowed. That region is not shown in Figure \ref{zskerr}. 
At any rate, in spiral galaxies, most of the stars have direct rather 
than retrogade orbits. Figure \ref{Masa05} presents the
mass parameter $M=M(r_c,z_1,z_2)$ 
for $r_c=1$ and $r_c= 3$.

\begin{figure}[htp]
  \centerline{ \epsfysize=7.5cm \epsfbox{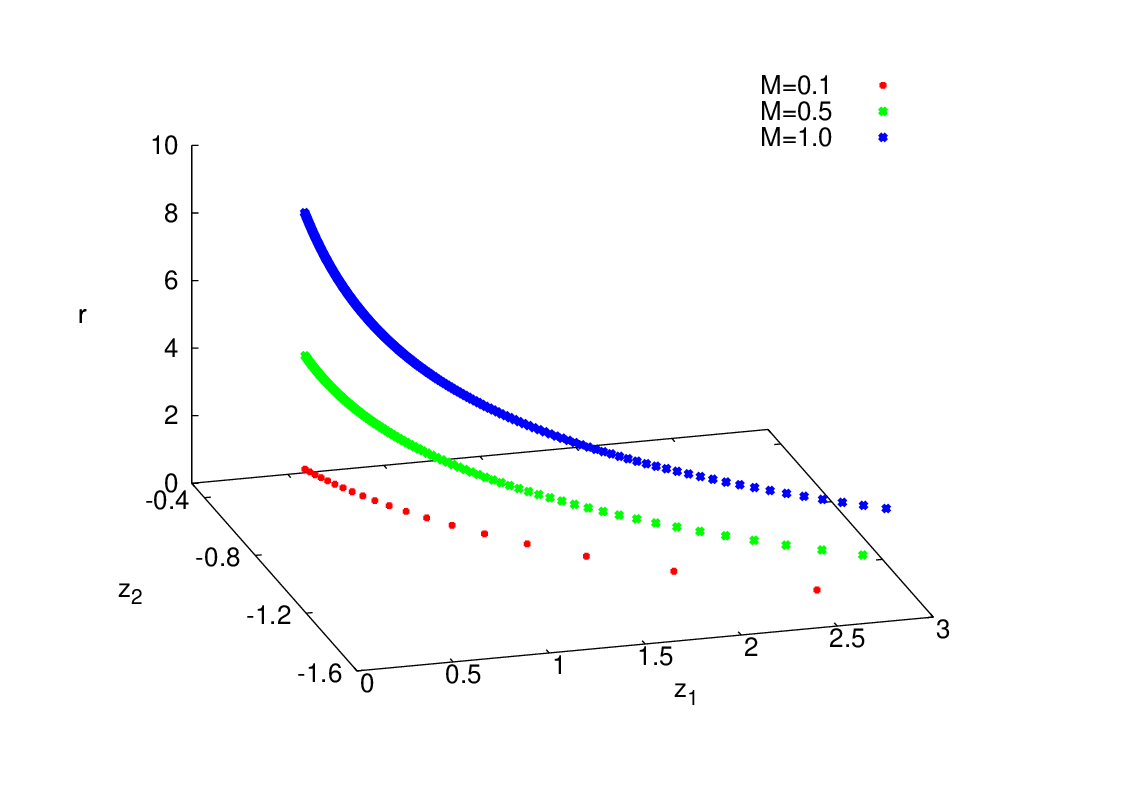}}
  \caption{For three different scaled mass parameters,
    the set of points $\{ (z_1 , z_2, r_c ) \}$ 
    corresponding to those values of $M$ is shown.  
    }
  \label{MsKerr}
\end{figure}

For some values of the mass parameter $M$, figure \ref{MsKerr} shows the
set of points $\{ (z_1 , z_2, r_c ) \}$ 
corresponding to those values of $M$. 
If a set of observations $\{ ( z_{red}, z_{blue}, r_c) \}$ 
of redshifts-blueshifts coming from emitters in circular orbits of radii 
$r_c$ laid along and around a curve corresponding to a value $M$, that specific 
value would be an estimate of the Kerr black hole mass $M$.

\begin{figure}[htp]
  \centerline{ \epsfysize=6cm \epsfbox{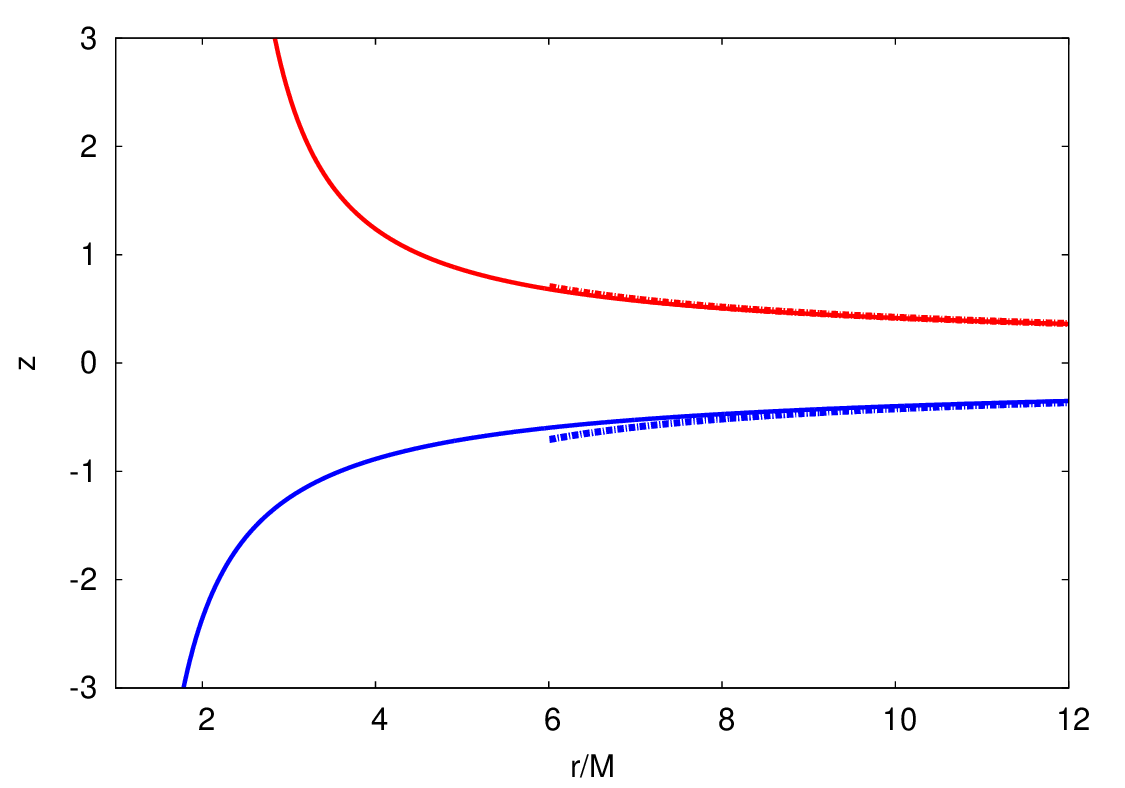}}
  \caption{We show $z_1$ (blueshift) and $z_2$ 
    (redshift) as a function of $r/M$ (solid blue 
    and red curve respectively),
    being $M$ the rotating black hole mass at the center of our galaxy.
    As $r/m$ increases $z_{red} \to -z_{blue}$ as is the case for 
    the Schwarzschild black hole (dashed curves).}
  \label{ZvsRKerr}
\end{figure}

If we select the estimate of the putative black hole
mass at the center of our galaxy $M = 2.72 \times 10^6 M_{\odot}$ 
to define $\tilde{r}=r/M$ and $a=pM=0.9939 M$, the expressions of the 
frequency shifts become

\begin{eqnarray}
z_1&=&\frac{\pm \left (2p+\tilde{r}\sqrt{\tilde{r}^2-2\tilde{r}+p^2}\right )}{\tilde{r}^{3/4}(\tilde{r}-2)\sqrt{\tilde{r}^{3/2}-3\tilde{r}^{1/2}\pm 2p} }
\nonumber \\
z_2&=&\frac{\pm \left (2p-\tilde{r}\sqrt{\tilde{r}^2-2\tilde{r}+p^2}\right )}{\tilde{r}^{3/4}(\tilde{r}-2)\sqrt{\tilde{r}^{3/2}-3\tilde{r}^{1/2}\pm 2p} },
\nonumber
\end{eqnarray}

\noindent whose plots are shown in figure \ref{ZvsRKerr} for the corotating 
case. As $r/M \to 2$, $z_1 \to \infty$. Negative values 
of $z_1$ are found for $r_c < 2$, that might be due
to the very strong dragging of the black hole over the emitter. 
As $r/M$ increases, $z_{red} \to -z_{blue}$ as is the case for the
Schwarzschild black hole, whose plot is also shown (dashed curves) and starts
at $r=6$ as it should be.\\

\section{Final remarks}

In this paper we have applied the theoretical approch developed by H-N
to determine the mass parameter of compact objects in terms of the frequency
shifts $z$ of light emitted by particles traveling along circular geodesics of
radii $r_c$ around those objects. For the Schwarschild and Reissner 
Nordstr\"om black holes, we found an explicit formula $M=M(z,r_c)$ 
and $M=M(z,r_c,Q^2)$ respectively, and bounds for $z$. 
Not all values of $z$ would be detected from a far 
away observer.
For Boson Stars, $z$ increases as the radius of the orbits increases
and reaches a maximum shown in figure \ref{shiftbosones}. 
For different  equilibrium 
configurations this $z_{max}$ increases as the central value $\phi_0$ 
increases regardless that configuration lays on the stable or unstable
branch. The curve $z(\phi_{crit})$ seems to be the limit of all $z(\phi)$
for large radii. For configurations with a fixed $M$ but different $\Lambda$,
$z_{max}$ decreases as $\Lambda$ increases.
It would be interesting to perform a similar analysis
for rotating boson stars, this work is progress.
 
For the Kerr black hole, the mass parameter obtained as a root of the
polynomial $F(N;r_c,z_1,z_2)$ is nearly unique. There is a small region
in the space $\mathcal{D}$ where there are double roots. The plot of
the redshift and blueshift as a function of $r_c$ for the putative black hole
at the center of our galaxy was also presented. 
Recently, a black hole with scalar hair was constructed by 
Carlos Herdeiro and Eugen Radu \cite{radu}.
It would be interesting to construct
the curve $z=z(r_c)$ for a given $M$ for such space-time and 
compare it with the one presented here for the Kerr black hole to 
determine the effect of hair. \\

\noindent {\bf Acknowledgments}\\

R. B. is grateful to professor Harry Swinney for his warm hospitality at
the Center for Nonlinear Dynamics of the University of Texas at Austin
where part of this work was carried out. 
U. N. and R. B. acknowledge partial support by CIC-UMSNH.
S. V. acknowledges support by CONACyT, under retention grant. 
The authors thank SNI and PRODEP-SEP for support.

\end{document}